\documentclass[fleqn,twoside]{article}
\usepackage[headings]{espcrc2}
\usepackage{graphicx}
\usepackage[figuresright]{rotating}
\usepackage{epsfig}

\title{Theoretical status of $B_s$-mixing and lifetimes of heavy hadrons}

\author{Alexander Lenz \address{
Institut f{\"u}r theoretische Physik,
Universit{\"a}t Regensburg, 
D-93040 Regensburg, Germany}}

\runtitle{Theoretical status of $B_s$-mixing and lifetimes of heavy hadrons}
\runauthor{A. Lenz}

\begin{document}

\begin{abstract}
We review the theoretical status of the lifetime ratios 
$\tau_{B^+} / \tau_{B_d}$, $\tau_{B_s} / \tau_{B_d}$, 
$\tau_{\Lambda_b} / \tau_{B_d}$  and $\tau_{B_c}$ and of the mixing 
quantities $\Delta M_s$, $\Delta \Gamma_s$ and $\phi_s$.
$\Delta M_s$ and  $\Delta \Gamma_s$ suffer from large uncertainties due to
the badly known decay constants, while the ratio 
$\Delta \Gamma_s / \Delta M_s$ can be determined 
with almost no non-perturbative uncertainties, therefore it can 
be used perfectly to find possible new physics contributions in 
the mixing parameters. We suggest a very clear method 
of visualizing the bounds on new physics and demonstrate this by 
combining the latest experimental numbers on the mixing quantities 
quantities with theory - one already gets some hints for new physics 
contributions, but more precise experimental numbers are needed to draw
some definite conclusions. We conclude with a ranking list of all the
discussed quantities according to their current theoretical uncertainties
and point out possible improvements.
\end{abstract}
 
\maketitle

\section{Introduction}

Inclusive decays (see e.g.\cite{inclusive} and references therein) 
and lifetimes of heavy mesons can be calculated within the framework of the
heavy quark expansion (HQE) \cite{HQE1}. In this approach
the decay rate is calculated in an expansion 
in inverse powers of the heavy b-quark mass:
$\Gamma = \Gamma_0 + \Lambda^2 / m_b^2  \, \, \Gamma_2
                   + \Lambda^3 / m_b^3  \, \, \Gamma_3 + \ldots $.
$\Gamma_0$ represents the decay of a free heavy b-quark, 
according to this contribution all b-mesons have the same lifetime.
The first correction arises at order $1/m_b^2$, they are due to the 
kinetic and the chromomagnetic operator. At order $1/m_b^3$ the 
spectator quark gets involved in the weak 
annihilation and Pauli interference diagrams \cite{HQE1,spectator}. 
This contributions are numerically enhanced by a phase space factor 
of $16 \pi^2$. Each of the $\Gamma_i$ contains perturbatively calculable 
Wilson coefficients and non-perturbative parameters, like decay constants or
bag parameters. 
This approach clearly has to be distinguished from QCD inspired models.
It is derived directly from QCD and the basic assumptions
(convergence of the expansion in $\alpha_s$ and $\Lambda / m_b$) can be 
simply tested by comparing experiment and theory for different quantities
(see e.g. \cite{updates}).
\vspace{-0.25cm}
\section{Lifetimes}
The lifetime ratio of two heavy mesons reads
\vspace{-0.25cm}
\begin{displaymath}
\frac{\tau_1}{\tau_2} = 1 +
\frac{\Lambda^3}{m_b^3} 
\left( \Gamma_3^{(0)} \! \!  + 
\frac{\alpha_s}{4 \pi} \Gamma_3^{(1)} \! \! + \ldots\right) +
\frac{\Lambda^4}{m_b^4} 
\left( \Gamma_4^{(0)} \! \! + \ldots\right) 
\end{displaymath}
If one neglects small isospin or SU(3) violating effects one 
has no $1/m_b^2$ corrections \footnote{In the case of 
$\tau_{\Lambda_b}/ \tau_{B_d}$ these effects are expected to be of the order
of $5\%$.} and a deviation of the lifetime ratio from
one starts at order $1/m_b^3$.
For the ratio $\tau_{B^+}/ \tau_{B_d}$ the leading term $\Gamma_3^{(0)}$ has been determined in \cite{HQE1,tauLO}.
For a quantitative treatment of the lifetime ratios NLO QCD corrections 
are mandatory -  $\Gamma_3^{(1)}$ has been determined in 
\cite{BBGLN02}. Subleading effects of ${\cal O} (1/m_b)$ turned 
out to be negligible \cite{lifetime1m}.
Updating the result from \cite{BBGLN02} with  matrix elements from
\cite{lifetimelattice}
and the values $V_{cb} = 0.0415$, $m_b = 4.63$ GeV and $f_{B} = 216 $ MeV
\cite{fB} we obtain a value,
which is in excellent agreement with the experimental number 
\cite{HFAG,HCP07EXP}:
\begin{displaymath}
\frac{\tau(B^+)}{\tau(B_d^0)}_{\rm NLO} \hspace{-0.7cm} = 
1.063\pm 0.027 ,
\frac{\tau(B^+)}{\tau(B_d^0)}_{\rm Exp} \hspace{-0.6cm} = 
1.076\pm 0.008 .
\end{displaymath}
To improve the theoretical accuracy further we need more precise 
lattice values, in particular of the appearing color-suppressed 
operators.
In the lifetime ratio $\tau_{B_s}/ \tau_{B_d}$ a cancellation of 
weak annihilation contributions arises, that differ only by small
SU(3)-violation effects. One expects a number that is close to one
\cite{tauLO,BBGLN02,tauBs,BBD}. The experimental number
\cite{HFAG,HCP07EXP} is slightly smaller
\begin{displaymath}
\frac{\tau(B_s)}{\tau(B_d)}_{\rm Theo}  \hspace{-0.3cm}= 1.00 \pm  0.01 ,
\frac{\tau(B_s)}{\tau(B_d)}_{\rm Exp}   \hspace{-0.3cm}= 0.950 \pm 0.019.
\end{displaymath}
Next we consider two hadrons, where the theoretical situation is much worse 
compared to the mesons discussed above.
The lifetime of the doubly heavy meson $B_c$ has been investigated in
\cite{taubctheory}, but only in LO QCD.
\begin{displaymath}
\tau(B_c)_{\rm LO} = 0.52_{-0.12}^{+0.18} \,  \mbox{ps},
\tau(B_c)_{\rm Exp}  = 0.460 \pm 0.066 \, \mbox{ps}
\end{displaymath}
In addition to the b-quark now also the c-charm quark can decay, giving rise
to the biggest contribution to the total decay rate.
The current experimental number is taken from \cite{taubcexp,HFAG}. 
In the case of the $\Lambda_b$-baryon 
the NLO-QCD corrections are not complete and there are only preliminary 
lattice studies  for a part of the arising matrix elements, see e.g. 
\cite{Ceciliatalk}, so the theoretical error has to be met with some
skepticism. Moreover there are some discrepancies in the experimental
numbers \cite{HFAG,Lambdalifeexp}.
\begin{displaymath}
\frac{\tau(\Lambda_b)}{\tau(B_d)}_{\rm Theo} \hspace{-0.6cm} = 0.88 \pm  0.05 \, \, \, \,,
\frac{\tau(\Lambda_b)}{\tau(B_d)}_{\rm Exp}  \hspace{-0.6cm} = 0.912 \pm 0.032\, .
\end{displaymath}

\section{Mixing Parameters}

\begin{figure}[htb]
\begin{center}
\epsfig{file=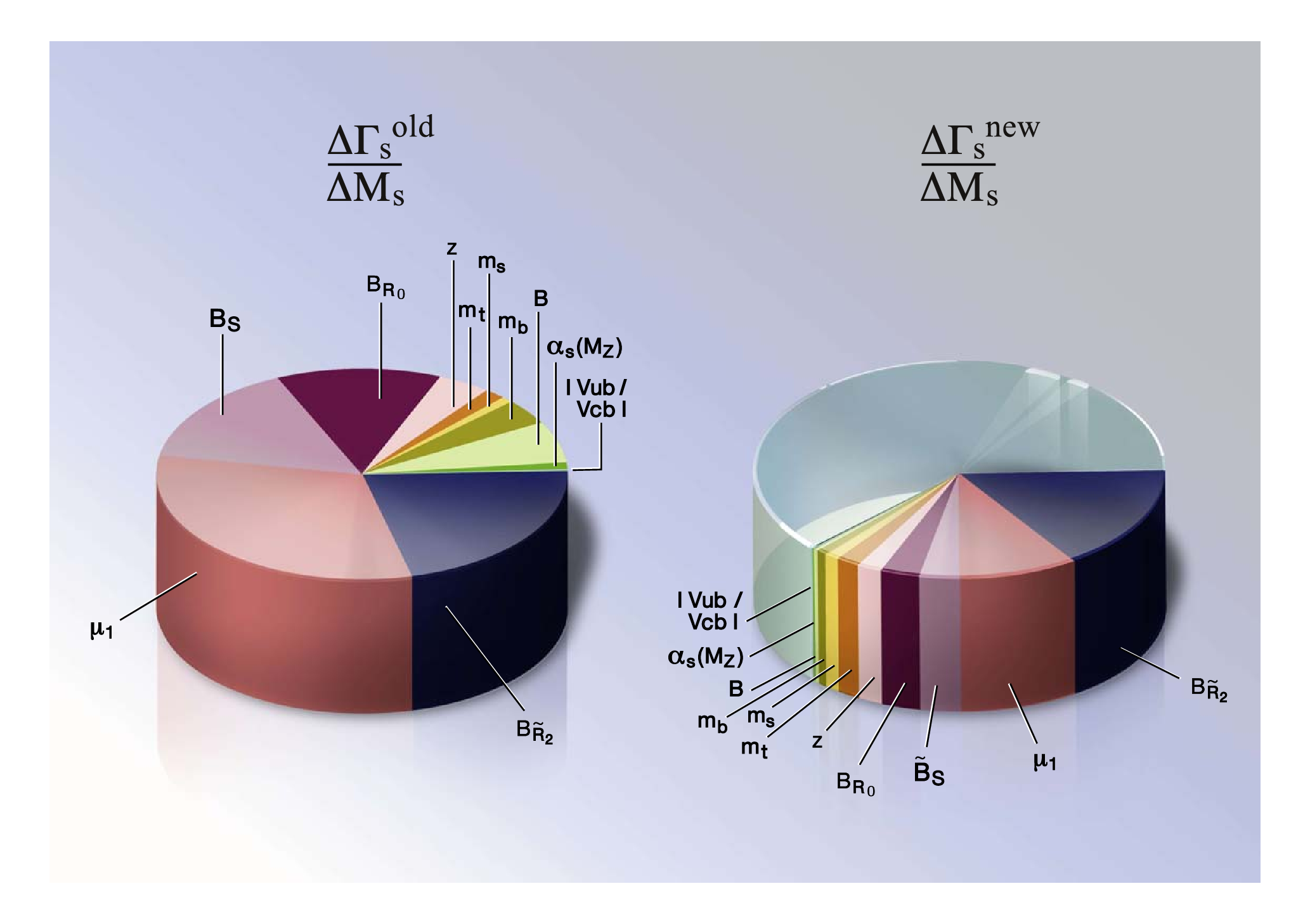,height=2.0 in}
\caption{Error budget for the theoretical determination of 
$\Delta \Gamma_s / \Delta M_s $. Compared to previous approaches (left)
the new strategy lead to a reduction of the theoretical error by almost 
a factor of 3.}
\label{fig:Kuchen}
\end{center}
\end{figure}

\begin{figure}
\includegraphics[width=0.45\textwidth,angle=0]{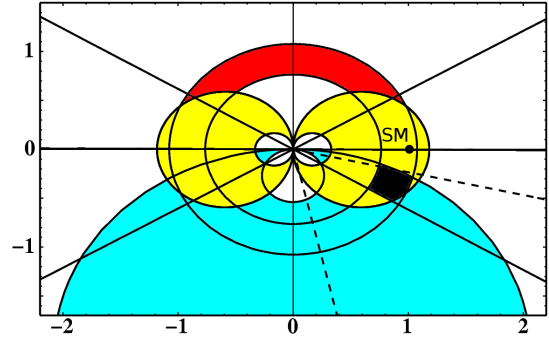}
\caption{Current experimental bounds in the complex $\Delta_s$-plane.
  The bound from $\Delta M_s$ is given by the red (dark-grey) ring around
  the origin. The bound from $\Delta \Gamma_s / \Delta M_s$ is given
  by the yellow (light-grey) region and the bound from $a_{fs}^s$ is given
  by the light-blue (grey) region. The angle $\phi_s^\Delta$ can be extracted
  from $\Delta \Gamma_s$ (solid lines) with a four fold ambiguity - one bound
  coincides with the x-axis! - or from the angular analysis in 
  $B_s \to J / \Psi \phi$ (dashed line). If the standard model is valid 
  all bounds should coincide in the point (1,0). The current experimental 
  situation shows a small deviation, which might become significant, if the 
  experimental uncertainties in $\Delta \Gamma_s$, $a_{sl}^s$ and $\phi_s$ 
  will go down in near future.}\label{boundbandreal}
\end{figure}

In this section we briefly investigate the status of the mixing parameters.
For a more detailed review we refer the interested reader to \cite{LN}.
\\
The mixing of the neutral B-mesons is described by the off diagonal
elements $\Gamma_{12}$ and $M_{12}$ of the mixing matrix.
$\Gamma_{12}$ stems from the absorptive part of the box diagrams - only
internal up and charm quarks contribute, while $M_{12}$ stems from
the dispersive part of the box diagram, therefore being sensitive to 
heavy internal  particles like the top quark or heavy new physics particles.
The calculable quantities {$  |M_{12}|$}, {$ |\Gamma_{12}|$} and 
{$ \phi = \mbox{arg}( -M_{12}/\Gamma_{12})$}
can be related to three observables (see \cite{LN,BBLN03} for a detailed 
description):
\begin{itemize}
\vspace{-0.2cm}
\item \mbox{Mass difference 
       $ \Delta M \approx 2 { |M_{12}|} $}
\vspace{-0.2cm}
\item \mbox{Decay rate difference 
      $ \Delta \Gamma \approx 
        2 { |\Gamma_{12}| \cos  \phi } $}
\vspace{-0.2cm}
\item Flavor specific or semi-leptonic CP asymmetries:
      $a_{fs} =  { \mbox{Im} \frac{\Gamma_{12}}{M_{12}}} =
      \frac{\Delta \Gamma}{\Delta M} \tan \phi $.
\end{itemize}
Calculating the box diagram with internal top quarks one obtains
 \begin{displaymath}
        M_{12,q} =  \frac{G_F^2}{12 \pi^2} 
          (V_{tq}^* V_{tb})^2 M_W^2 S_0(x_t)
          {B_{B_q} f_{B_q}^2  M_{B_q}} \hat{\eta }_B
        \end{displaymath}
The Inami-Lim function $S_0 (x_t = \bar{m}_t^2/M_W^2)$ 
\cite{IL} is the result of the box diagram 
without any gluon corrections. The NLO QCD correction is parameterized by 
$\hat{\eta}_B \approx 0.84$ \cite{BJW}.
The non-perturbative matrix element is parameterized by the 
bag parameter $B$ and the decay constant $f_B$.
Using the conservative estimate $f_{B_s} = 240 \pm 40$ MeV \cite{LN} 
and the bag parameter $B$ from JLQCD \cite{JLQCD} we obtain in units of
$\mbox{ps}^{-1}$ (experiment from \cite{HFAG,HCP07EXP,deltamsexp})
\begin{displaymath}
\Delta M_s^{\rm Theo} \hspace{-0.2cm}= 19.3 \pm 6.4 \pm 1.9,
\Delta M_s^{\rm Exp}  \hspace{-0.2cm}=  17.77 \pm 0.12  
\end{displaymath}
The first error in the theory prediction stems from the uncertainty in 
$f_{B_s}$ and the second error
summarizes the remaining theoretical uncertainties.
The determination of $\Delta M_d$ is affected by even larger 
uncertainties because here one has to extrapolate the decay constant 
to the small mass
of the down-quark.
The ratio $\Delta M_s / \Delta M_d$ is theoretically better under
control since in the ratio of the non-perturbative parameters many systematic
errors cancel, but on the other hand it is affected by large uncertainties
due to $|V_{ts}|^2/|V_{td}|^2$.
To be able to distinguish possible new physics contributions to $\Delta M_s$
from QCD uncertainties much more precise numbers for $f_{B_s}$ are needed.
\\
In order to determine the decay rate difference of the neutral B-mesons 
and flavor specific CP asymmetries a precise determination of $\Gamma_{12}$ 
is needed, which can be written as
\begin{displaymath}
\Gamma_{12} = 
\frac{\Lambda^3}{m_b^3} \left( \Gamma_3^{(0)} + \frac{\alpha_s}{4 \pi} \Gamma_3^{(1)} + \ldots\right) +
\frac{\Lambda^4}{m_b^4} \left( \Gamma_4^{(0)} + \ldots\right)
\end{displaymath}
The leading term $\Gamma_3^{(0)}$ was determined in \cite{dgLO}.
The numerical and conceptual important NLO-QCD corrections ($\Gamma_3^{(1)}$) 
were determined in \cite{BBGLN98,BBLN03}.
Subleading $1/m$-corrections, i.e. $\Gamma_4^{(0)}$ were calculated 
in \cite{BBD,1overm}
and even the Wilson coefficients of the $1/m^2$-corrections 
($\Gamma_5^{(0)}$) were calculated and found to be small \cite{LN2}.
In \cite{LN} a strategy was worked out to reduce the theoretical uncertainty
in $\Gamma_{12}/M_{12}$ by almost a factor of 3, see Fig. (\ref{fig:Kuchen}) 
for an illustration.
One gets
\begin{displaymath}
\frac{\Delta \Gamma_s}{\Delta M_s} = 
10^{-4} \cdot
\left[ 46.2  + 10.6 \frac{ B_S'}{B}  - 11.9  \frac{B_R}{B} 
\right] 
\end{displaymath}
The dominant part of $\Delta \Gamma / \Delta M $ can now be determined without 
any hadronic uncertainties (for more details see \cite{LN})!
Using the non-perturbative parameters from
\cite{JLQCD,Becirevic}, we obtain the following final numbers
(see \cite{LN} for the complete list of the numerical values of the 
input parameters and the very conservative ranges in which we varied them)
\begin{eqnarray}
\Delta \Gamma_s & \hspace{-0.6cm}= & \hspace{-0.6cm}
\left( 0.096   \pm 0.039 \right) \mbox{ps}^{-1},
\frac{\Delta \Gamma_s}{\Gamma_s}  
= 0.147 \pm 0.060,
\nonumber
\\
a_{fs}^s & \hspace{-0.6cm}= & \hspace{-0.6cm} 
\left( 2.06 \pm 0.57 \right) \cdot 10^{-5},
\frac{\Delta \Gamma_s}{\Delta M_s}  =  
\left( 49. 7 \pm 9.4 \right) 
 10^{-4} \! \! \! \! \! ,
\nonumber
\\
\phi_s &  \hspace{-0.1cm}= & 
0.0041 \pm 0.0008 \, \, \, = \, \, \, 0.24^\circ \pm 0.04 \, .
\nonumber
\end{eqnarray}
New physics (see e.g. references in \cite{LN})
is expected to have almost no impact on $\Gamma_{12}$, but it 
can change $M_{12}$ considerably -- we denote the deviation factor by the 
complex number $\Delta$. Therefore one can write
\begin{displaymath}
\Gamma_{12,s} =  \Gamma_{12,s}^{\rm SM},
M_{12,s}  =  M_{12,s}^{\rm SM} \cdot { \Delta_s};
{ \Delta_s} = { |\Delta_s|} e^{i  \phi^\Delta_s} 
\end{displaymath}
With this parameterisation the physical mixing parameters can be written as
\begin{eqnarray}
 \Delta M_s  & =  & 2 | M_{12,s}^{\rm SM} | \cdot { |\Delta_s |} 
\label{bounddm},
\nonumber
\\
\Delta \Gamma_s   & =  &2 |\Gamma_{12,s}|
\cdot \cos \left( \phi_s^{\rm SM} + { \phi^\Delta_s} \right),
\label{bounddg}
\nonumber
\\
\frac{\Delta \Gamma_s}{\Delta M_s} 
& = &
 \frac{|\Gamma_{12,s}|}{|M_{12,s}^{\rm SM}|} 
\cdot \frac{\cos \left( \phi_s^{\rm SM} + { \phi^\Delta_s} \right)}
{ |\Delta_s|}
\label{bounddgdm},
\nonumber
\\
a_{fs}^s 
& = &
 \frac{|\Gamma_{12,s}|}{|M_{12,s}^{\rm SM}|} 
\cdot \frac{\sin \left( \phi_s^{\rm SM} + { \phi^\Delta_s} \right)}
{ |\Delta_s|}.
\label{boundafs}
\end{eqnarray}
Note that $\Gamma_{12,s} / M_{12,s}^{\rm SM}$ is now due to the improvements in
\cite{LN} theoretically very well under control.
Next we combine the current experimental numbers with the theoretical 
predictions to extract bounds in the imaginary $\Delta_s$-plane by the use of
Eqs. (\ref{boundafs}), see Fig. (\ref{boundbandreal}). 
The width difference $\Delta \Gamma_s /\Gamma_s $ was 
investigated in \cite{dgexp,HCP07EXP}. 
The semi-leptonic CP asymmetry in the $B_s$ system has been determined 
in \cite{HCP07EXP,aslsexp} (see \cite{LN} for more details).
Therefore we use as experimental input
\begin{eqnarray}
\Delta \Gamma_s =  0.17 \pm 0.09 \, \mbox{ps}^{-1} ,
&& \hspace{-0.6cm}
\phi_s  =  -0.79  \pm 0.56 .
\nonumber
\\
a_{sl}^{s} = \left(- 5.2 \pm 3.9 \right) \cdot 10^{-3} \, .
&&
\nonumber
\end{eqnarray}

\section{Conclusion and outlook} 
We have reviewed the theoretical status of lifetimes of heavy hadrons
and the measureable mixing quantities of the neutral B-mesons. Both classes of 
quantities can be described with the help of the HQE - a systematic expansion
based simply on QCD.
Inspired by the current theoretical uncertainties we put these quantities
in 3 classes
\begin{itemize}
\item[1a:] Gold:   $\tau (B_s) / \tau (B_d)$, $\phi_q$, $a_{fs}^q$
\item[1b:] Silver: $\tau (B^+) / \tau (B_d)$, $\Delta \Gamma / \Delta M$  
\item[2:]  Iron:  $\Delta \Gamma$, $\Delta M$, $\Delta M_s / \Delta M_d$,
             $\tau (B_c)$, $\tau (\Lambda_b)$
\end{itemize}
{\it Gold} and {\it silver} have similar theoretical uncertainties, but 
the {\it gold}-quantities are predicted to be small numbers, so a sizeable 
experimental result corresponds unambigously to some new effect.

Let's start with the worst.
Iron: The theoretical uncertainty in the mixing parameters $\Delta M$ and 
$\Delta \Gamma$ is completely dominated by the decay  constant.
Here some progress on the non-perturbative side is mandatory.
In $\Delta M_s / \Delta M_d$ the dominant uncertainty is given by 
$|V_{ts}/V_{td}|^2$. In $\tau (B_c)$ and $\tau (\Lambda_b)$ the important
NLO-QCD are missing or are incomplete, moreover we have only preliminary 
lattice studies of the non-perturbative matrix elements.
\\
Silver:
Theoretical predictions of $\tau_{B^+}/\tau_{B_d}$ are in excellent 
agreement with the experimental numbers. We do not see any signal of possible 
duality violations. To become even more 
quantitative in the prediction of $\tau_{B^+}/\tau_{B_d}$ 
the non-perturbative estimates of the bag parameters - in particular of the
color-suppressed ones - have to be improved.
In \cite{LN} a method was worked out  to reduce the 
theoretical error in $\Delta \Gamma / \Delta M$  considerably.
For a further reduction of the theoretical uncertainty
in the mixing quantities the unknown matrix elements of the power 
suppressed operators  have to be determined.
Here any non-perturbative estimate would be very desirable.
A first step in that direction was performed in \cite{Siegen}.
If accurate non-perturbative parameters are available one might think about 
NNLO calculations ($ \alpha_s/m_b$- or $\alpha_s^2$-corrections) to
reduce the remaining $\mu$-dependence and the uncertainties due to the 
missing definition of the b-quark mass in the power corrections.
\\
Gold: The improvements for $\Delta \Gamma / \Delta M$ apply to
$a_{fs}$ and $\Phi_q$ as well.

The relatively clean standard model predictions for the mixing
quantities can be used to look for 
new physics effects in $B_s$-mixing. From the currently available experimental 
bounds on $\Delta \Gamma_s$ and $a_{fs}$ one already gets some hints for
deviations from the standard model.
To settle this issue we are eagerly waiting for more data from TeVatron!

I would like to thank the organizers and convenors of HCP2007 for the 
invitation and the financial support, which enabled me to take part in this
wonderful symposium and 
Uli Nierste for the pleasant collaboration.

\underline{\bf Note added:}
There is sometimes a confusion between the mixing phases $\beta_s$ 
and $\phi_s$, which we would like to adress here. Both numbers are 
expected to be small in the standard model -
$\phi_s =  (0.24 \pm 0.04)^\circ$ and 
$ 2 \beta_s =  (2.2 \pm 0.6)^\circ (= (0.04 \pm 0.01) \mbox{rad} )$, 
but in view of the high future experimental precisions - in particular at
LHCb \cite{LHCb}- a clear distinction 
might be useful.
\\
$ 2 \beta_s := - \mbox{arg} [(V_{tb}V_{ts}^*)^2 /(V_{cb}V_{cs}^*)^2 ]$
is the phase which appears in $b \to c \bar{c} s$ decays of neutral
B-mesons taking possible mixing into account, 
so e.g. in the case $B_s \to J/\psi + \phi$.
$(V_{tb}V_{ts}^*)^2$ comes from the mixing (due to $M_{12}$) and 
$(V_{cb}V_{cs}^*)^2$ comes from the ratio of $b \to c \bar{c} s$ decay 
and $\bar{b} \to \bar{c} c \bar{s}$ amplitudes.
Sometimes $ \beta_s $ is approximated as 
$ 2 \beta_s \approx  - \mbox{arg} [(V_{tb}V_{ts}^*)^2] \approx 
- \mbox{arg} [(V_{ts}^*)^2] $ - the error due to this approximation is on the
per mille level.
\\
$\phi_s := \mbox{arg} [-M_{12}/\Gamma_{12}]$ is the phase that appears e.g. in
$a_{fs}^s$. In $M_{12}$ we have again $(V_{tb}V_{ts}^*)^2$, while we have a 
linear combination of  $(V_{cb}V_{cs}^*)^2$, $V_{cb}V_{cs}^*V_{ub}V_{us}^*$ 
and $(V_{ub}V_{us}^*)^2$ in $\Gamma_{12}$. Neglecting the latter two 
contributions - which is not justified - would yield the phase $2 \beta_s$.

New physics alters the phase $-2 \beta_s$ to $\phi_s^\Delta-2 \beta_s$
and the phase $\phi_s$ to $\phi_s^\Delta +  \phi_s$.
If the new physics contribution is sizeable, then in both cases only
$\phi_s^\Delta$ survives, since the standard model phases are very small.

\end{document}